\documentclass[aps,prstab,groupedaddress,floatfix,byrevtex,twoside,superscriptaddress,preprint]{revtex4}
\ifx\pdftexversion\undefined
  \usepackage[dvips]{graphicx}
\else
  \usepackage[pdftex]{graphicx}
\fi

\usepackage{color}
\usepackage{amsmath}
\usepackage[psamsfonts]{amssymb}
\usepackage{natbib}
\usepackage{hyperref}
\usepackage{pifont}
\usepackage{latexsym}
\setkeys{Gin}{width=\linewidth}

\begin{document}


\title{The RFOFO Ionization Cooling Ring for Muons}
\author{R.~Palmer}  \affiliation{Brookhaven National Laboratory, Upton, NY 11973, USA}
\author{V.~Balbekov} \affiliation{Fermi National Accelerator Laboratory,
Batavia, IL 60510, USA}
\author{J.S.~Berg} \affiliation{Brookhaven National Laboratory, Upton, NY 11973, USA}
\author{S.~Bracker}\affiliation{University of Mississippi, Oxford, MS
38677, USA}%
\author{L.~Cremaldi}\affiliation{University of Mississippi, Oxford, MS
38677, USA}%
\author{R.C.~Fernow} \affiliation{Brookhaven National Laboratory, Upton, NY 11973, USA}
\author{J.C.~Gallardo} \affiliation{Brookhaven National Laboratory, Upton, NY 11973, USA}
\author{R.~Godang}\affiliation{University of Mississippi, Oxford, MS
38677, USA}%
\author{G.~Hanson}\affiliation{University of California, Riverside, CA 
92521, USA}%
\author{A.~Klier}\affiliation{University of California, Riverside, CA 
92521, USA}%
\author{D.~Summers}\affiliation{University of Mississippi, Oxford, MS
38677, USA}%

\date{\today}

\begin{abstract}
Practical ionization cooling rings could lead to lower cost or improved 
performance in neutrino factory or muon collider designs. The ring modeled 
here uses realistic three-dimensional fields. The performance of the ring
compares favorably with the linear cooling channel used in the second US 
Neutrino Factory Study. The normalized 6D emittance of an ideal ring is 
decreased by a factor of approximately 240, compared with a factor of only 
15 for the linear channel. We also examine such \textit{real-world} effects
as windows on the absorbers and rf cavities and leaving empty lattice cells
 for injection and extraction. For realistic conditions the ring decreases
 the normalized 6D emittance by a factor of 49.
\end{abstract}
\maketitle


\section{Introduction \label{sec1}}
Designs for neutrino factories~\cite{status_report} and muon colliders~\cite{rfofo-ref2}
use ionization cooling to reduce the emittance of the muon beam prior to
acceleration. Ionization cooling is currently the only feasible option for
cooling the beam within the muon lifetime $(\tau_o = 2.19\,\mu\text{s}).$ If
muons alternately pass through a material absorber, and are then
re-accelerated, and if there is sufficient focusing at the absorber, then
the transverse phase space is reduced, i.e. the muons are cooled in the
transverse dimension. 

A consequence of the transverse cooling is an increase of the longitudinal 
phase space caused by the unfavorable slope of the dE/dx curve for momenta 
below the ionization minimum and by energy straggling in the material. The 
momentum spread can be reduced if dispersion is introduced and a
wedge-shaped absorber is placed such that high momentum particles pass
through more material than low momentum particles. However, when this 
procedure is carried out the beam width is increased. The process is thus 
primarily an exchange of emittance between the longitudinal and transverse 
dimensions. Moreover, when combined with transverse cooling in the material, all 
three dimensions can be cooled.

Early muon collider studies had assumed that transverse cooling and emittance
exchange would be done in alternating stages. The transverse cooling used 
straight channels with rf and absorbers, while the emittance exchange
employed bent solenoids and wedges. A serious problem with this approach
was found to be the necessary matching between the two types of lattices. 
However, there has been considerable progress over the past few years in 
achieving 6D ionization cooling in cooling rings~\cite{rfofo-ref3}. Ring coolers
can overcome much of this matching problem by rapidly alternating the two 
functions. 

The first ring cooler design~\cite{rfofo-ref4,rfofo-ref4a} was based on solenoid 
focusing. Alternate cells contained
\begin{itemize}
\item transverse cooling in a long solenoid containing acceleration and a
  single hydrogen absorber
\item emittance exchange in a cell containing two bending magnets, two opposed
  solenoids, and a LiH wedge.
\end{itemize}
 Matching between these cells has currently been 
achieved only with hard-edged magnetic fields. Each pair of cells is long 
enough that at least one half-integer betatron resonance is present within 
the momentum acceptance. 

Later quadrupole-focused rings~\cite{rfofo-ref5a,rfofo-ref5} took the process a step 
further, using a single cell type in which a wedge absorber cooled in both
 longitudinal and transverse phase space. These ideas eased the lattice
 design, but introduced the possibility for transverse emittance growth
 from energy straggling in the absorber, since it was now in a dispersive
 location. In addition, the weaker focusing in such quadrupole lattices
 limits the available amount of cooling and momentum acceptance. Current
 versions of this ring use edge-focused dipoles instead of quadrupoles, and
 high pressure gas instead of wedge absorbers.~\cite{rfofo-xxx}

\begin{figure}[!tpbh]
\begin{center}
\includegraphics[width=4.in]{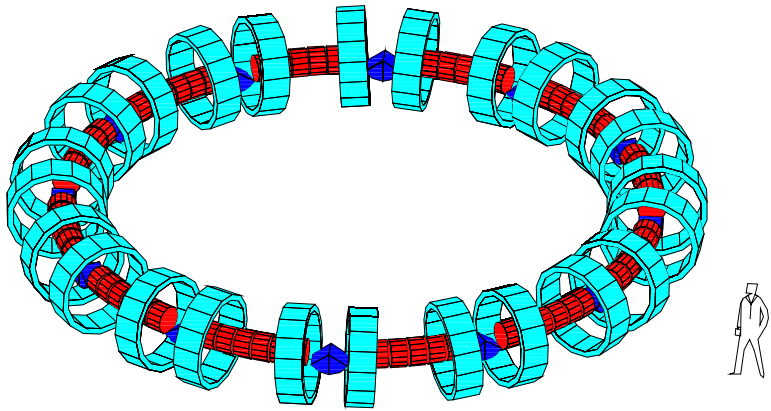} 
\end{center}
\caption{(Color) GEANT drawing of an idealized RFOFO cooling
  ring~\cite{rfofo-ref10}. The large cyan cylinders are solenoids, the
  small red cylinders are the active volume of the rf cavities, and the blue wedges are hydrogen absorbers.} 
\label{bfig1}
\end{figure}

  At present the most realistic modeling of a cooling ring has been done
  for the RFOFO cooling ring~\cite{rfofo-ref6,rfofo-ref7,rfofo-ref8,rfofo-ref9}. The FOFO part of the name refers
   to the
  focusing-drift-focusing nature of the solenoid lattice, in analogy to the
  FODO lattice for quadrupole channels. The prefix R designates a
  particular type of FOFO lattice where the magnetic cell has the same
  length as the geometric cell. As a result the axial field changes
  polarity in the middle of the cell. The ring discussed here follows the
  quadrupole design in the sense that it employs a single cell for doing both 
transverse
  cooling and emittance exchange. However, we use solenoidal focusing to
  obtain larger angular and momentum acceptances. The cell includes
  dispersion, acceleration, and energy loss in a single thick hydrogen
  wedge.  The overall layout of the ring is shown in Fig.~\ref{bfig1}.

Two solenoids in each cell with opposite polarity provide transverse
focusing. The solenoids are not equally spaced; those on either side of the
absorbers are closer together in order to increase the focusing at the
absorber. The vertical bending field is provided by alternately tipping the
 axes of
the solenoids above and below the orbital mid-plane. The bending also
provides the dispersion necessary for emittance exchange. A short cell
length is used to obtain a small beta function with a reasonable value of
the solenoid field strength. Wedge-shaped absorbers are placed in the beam
path at the locations where the solenoidal field changes direction and the
beta function is at a minimum.  The 
lattice has dispersion at the rf cavities, which introduces
synchro-betatron mixing and introduces some additional emittance growth. 

However, we find that the disadvantage of having dispersion at the rf
cavities is compensated by the greater
acceptance from the use of a single repeating cell with no integer or
half-integer betatron resonances in the momentum acceptance. In addition
the longitudinal cooling provided by the wedge-shaped absorber minimizes
losses from particles falling out of the rf bucket. Most of the
lattice cell is filled with rf cavities that restore the energy lost in the
absorbers, but empty cells are included where injection and extraction can
take place.

\section{Modeling the Ring \label{sec2}}
\begin{figure}[!tbhp] 
\begin{center} 
a)
\includegraphics[width=3.in]{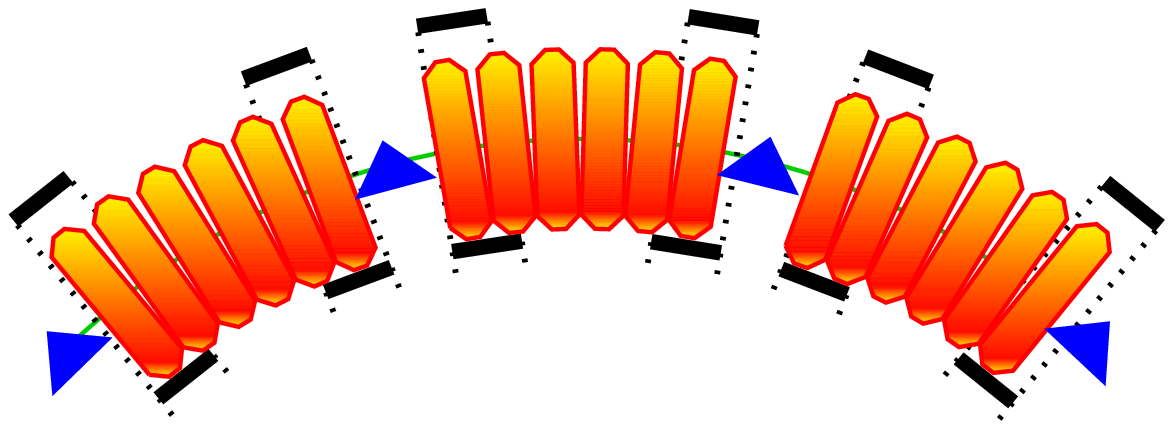}            

b)
\includegraphics[width=3.in]{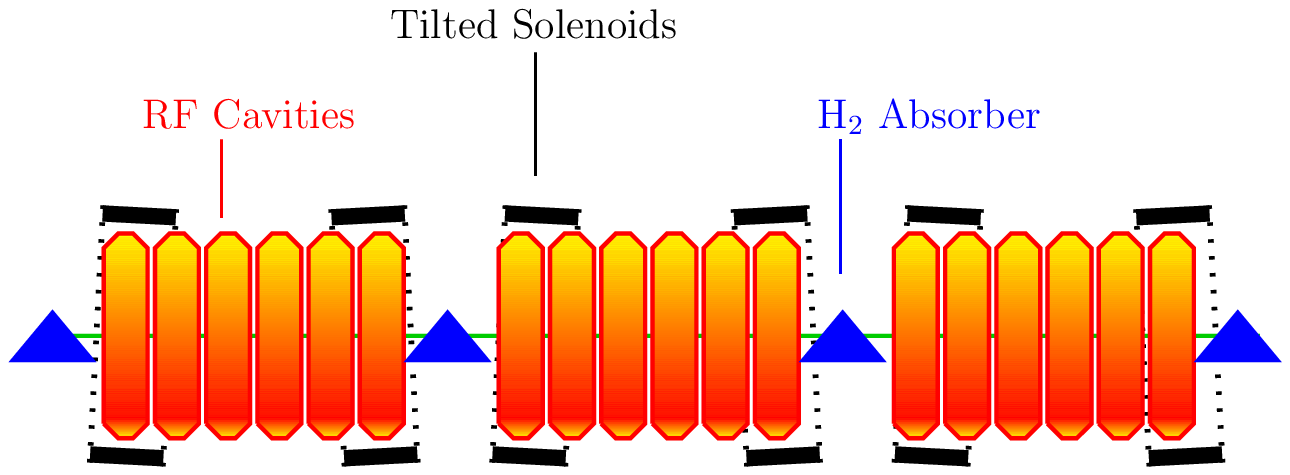}            
\end{center}
\caption{(Color) Three cells of the RFOFO lattice; a) plan view; b) side view. Notice that the coils have been displaced radially by 10~cm.} 
\label{bfig2}
\end{figure}

  A more detailed layout of three cells of the ring is shown in Fig.~\ref{bfig2} and a summary of ring properties is given in Table~\ref{tb1}.  

\begin{table}[!bthp]
\begin{center}
\caption{RFOFO ring parameters}
\label{tb1}
\begin{ruledtabular}
\begin{tabular}{lc}
Circumference (m) & 33\\
Total number of cells &12\\
Cells with rf cavities & 10\\
Maximum axial field (T) &2.77\\
Coil tilt angle (degree)&3\\
Average vertical field (T)&0.125\\
Average momentum (MeV/c)&220\\
Minimum transverse beta function (cm)&38\\
Maximum dispersion function (cm)&8\\
Wedge opening angle (degree)&100\\
Wedge thickness on-axis (cm)&28\\
Cavities rf frequency (MHz)&201.25\\
Peak rf gradient (MV/m)&12\\
Cavities rf phase from 0-crossing (degree)&25
\end{tabular}
\end{ruledtabular}
\end{center}
\end{table}
   The 33~m circumference ring is made up of 12~identical, 2.75~m long
   cells. The magnetic field is produced by solenoids of 50~cm
   in length, with an
   inner radius of 77~cm and an outer radius of 88~cm and a current density
   of $\pm 95.27\,\text{A/mm}^2.$ The longitudinal field on-axis has an
   approximately sinusoidal dependence on position. The tilt angle of the
   solenoids was adjusted to provide a mean vertical bending field of
   0.125~T. The centers of the solenoids are displaced radially outward
   from the 33~m reference circle by 10~cm to minimize the integrated
   on-axis radial field, which causes vertical beam deviations.

The RFOFO ring was modeled using three independent simulation codes. One private code was specifically written to examine this problem~\cite{rfofo-ref11,rfofo-ref12,rfofo-ref12a}.

The second code was ICOOL~\cite{icool}, which has been widely used for
neutrino factory and muon collider simulations. It includes a flexible
system for specifying the problem geometry and magnetic fields. For this
application the fields were either read in from an external file or
calculated from on-axis multipoles. RF cavities were modeled as pillboxes,
including the Bessel function radial dependence. Many of the energy loss,
scattering and straggling routines in ICOOL are based on \textbf{GEANT-3.21}~\cite{rfofo-ref14}.

The third code was the simulation software, MUC\_GEANT~\cite{rfofo-ref15}, which
is an application of \textbf{GEANT-3.21}, specially designed for muon cooling
simulations. The Runge-Kutta routine was changed to include electric
fields, so that rf acceleration would be simulated properly. The code 
is also data-driven, i.e. it allows the user to change the cooling channel
parameters, such as geometry, rf frequency/gradient and physics processes,
without recompiling. The magnetic field was read in from an
external file. The rf was modeled as a purely 
 axial field with no transverse variation.

The 3-dimensional magnetic field from the ring of tipped solenoids was
calculated in three other independent codes by summing the fields from a
set of current sheets~\cite{rfofo-ref12,rfofo-ref12a,rfofo-ref16,rfofo-ref17}. There are some
complexities in generating the fields for small rings which are much less
important for straight cooling channels. The coils being used in this
design are large-diameter, short, and carry high current; the maximum field
is several Tesla. The ring is small, a little over 5~m in radius. We
ignore for the moment the presence of any shielding iron. Then from every
point in the active (muon-accessible) region of the ring, a particle is
significantly affected by the field generated by all the coils, even the
ones on the other side of the ring. Hence one must develop a single global
field map, rather than assuming that a particle sees field components only
from the closest coils. 

\begin{figure}[!tbhp] 
\begin{center} 
\includegraphics*[width=0.4\linewidth]{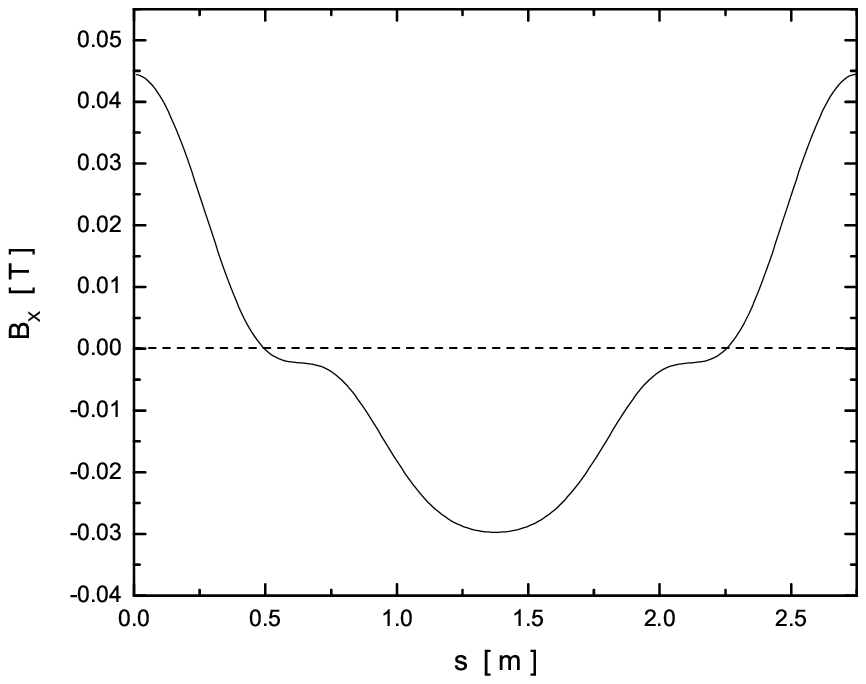}
\includegraphics*[width=0.4\linewidth]{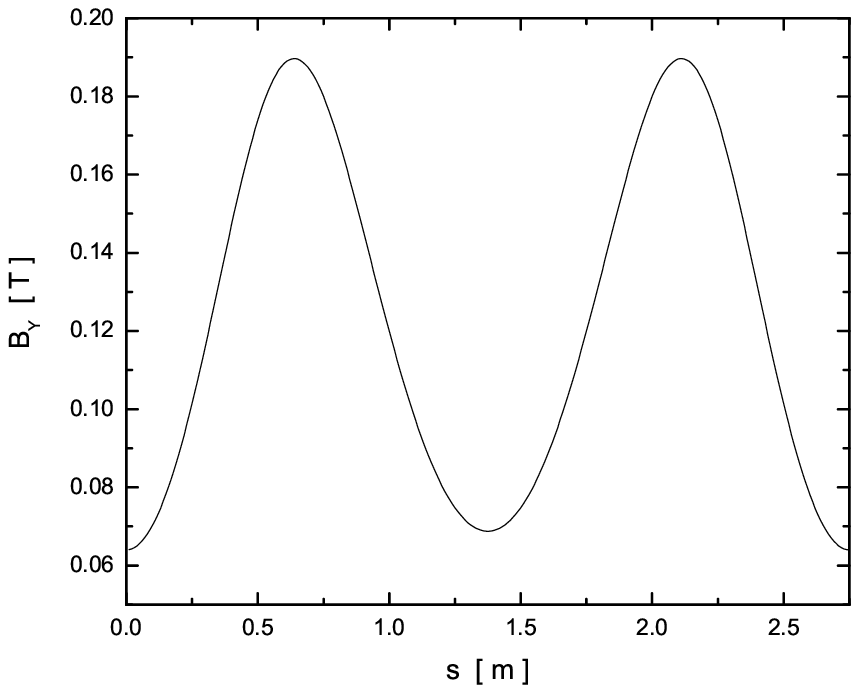}
\includegraphics*[width=0.4\linewidth]{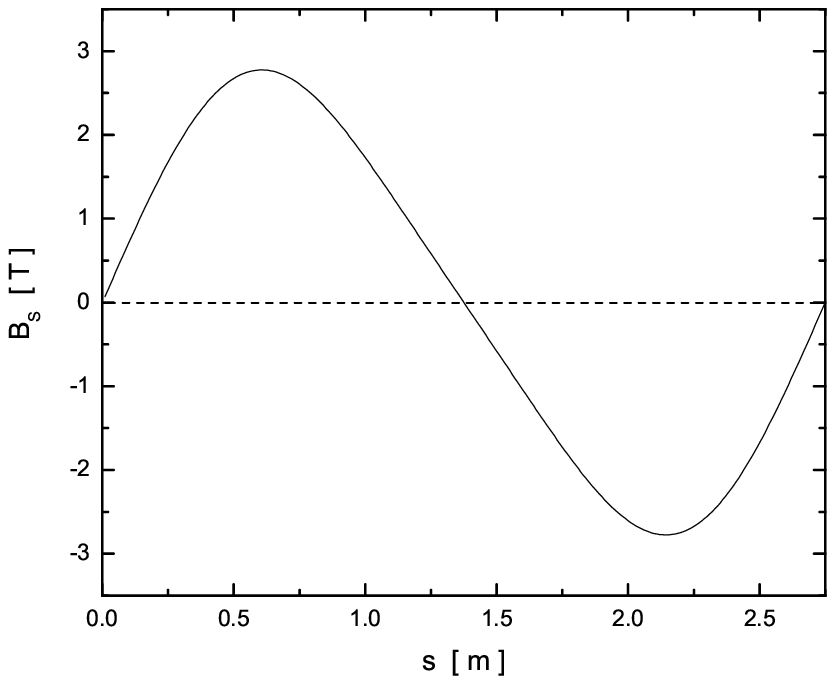}
\end{center}
\caption{Magnetic field on-axis for one cell. The upper panel are: the
  radial field $B_X,$ (left); the vertical field $B_Y$ (right) and the lower
  panel is the longitudinal field $B_S.$} 
\label{bfig3}
\end{figure}

The field of a solenoidal current sheet can be
written analytically in terms of elliptic integrals. The fields could also
be generated from a set of coils using the Biot-Savart law. The resultant
field components between the two methods agreed well and were shown to
satisfy Maxwell's equations to a high level of accuracy (approximately
$2~\mu$T/m). For GEANT a 3D field map was generated on a cartesian grid with
 1~cm spacing in each dimension. For ICOOL we used a 3D field map in
 cylindrical coordinates, also with 1~cm grid spacing in each dimension. Because the fields were modeled from 
coils, all magnetic end field
effects are automatically included in the simulations. The on-axis field 
components for one cell are shown in Fig.~\ref{bfig3}.

\begin{figure}[!tbhp] 
\begin{center} 
\includegraphics*[width=3.in]{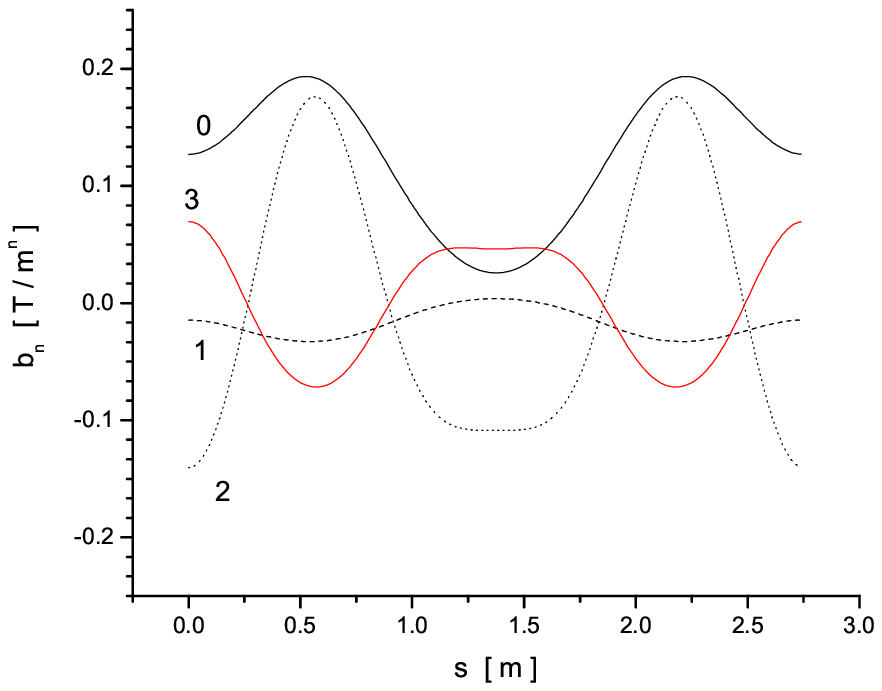}
\includegraphics*[width=3.in]{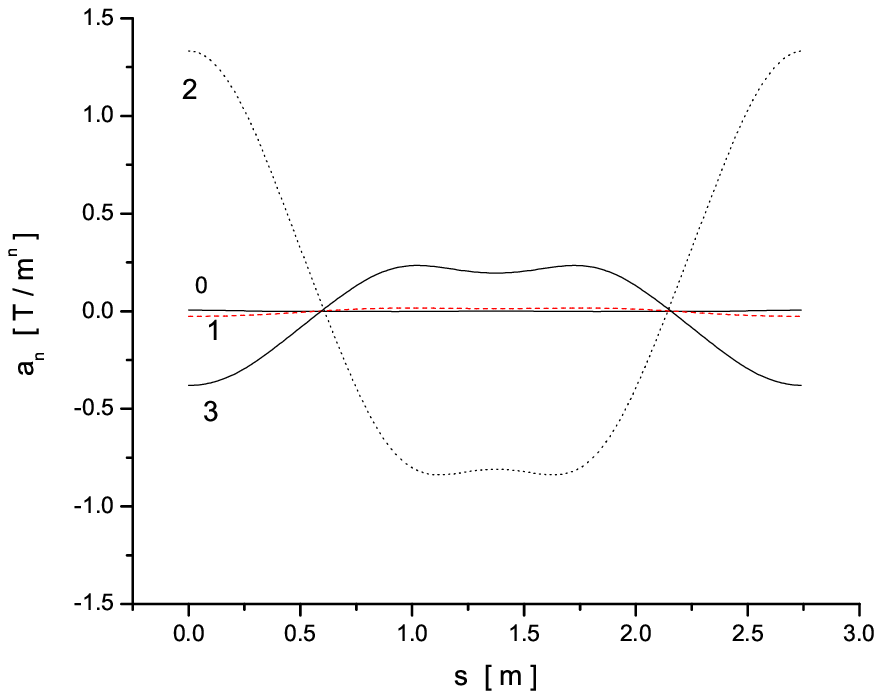}
\end{center}
\caption{(Color) On-axis normal (left) and skew (right) multipoles for one
  cell of the RFOFO ring. The curves are labelled with the multipole order,
$0$ for dipole, 1 for quadrupole, etc.} 
\label{bfig4}
\end{figure}

Once the magnetic field generator was working subsequent tracking could be
done from the field calculated on a grid and saved to a map, or from
multipoles~\cite{rfofo-ref18} of the field calculated around a reference
circle. Figure~\ref{bfig4} shows the calculated multipoles for one cell of 
the ring.
 The on-axis transverse field has a rich spectrum of higher normal and skew multipoles.

Each lattice cell contains 6~rf cavity cells, each 33~cm long. The transit
time factor for the cavity cells is 0.901. In ICOOL the rf cavities were
modeled using cylindrical pillboxes running in the TM$_{010}$ mode. The
frequency is 201~MHz and the peak accelerating gradient is 12~MV/m. The
cavity operates at a synchronous phase $25^{\circ}$ off the 0-crossing
point. The cavities are located in dipole fields. Since we are accelerating
muons, the cavities can be enclosed with metallic end windows in order to
produce the maximum electric field on-axis for a given amount of rf power
(maximum shunt impedance). The rf windows were stepped in thickness
radially, in order to provide minimum thickness near the beam axis and
still control the temperature increase due to rf heating.
  
\begin{figure}[!tbhp] 
\begin{center} 
\includegraphics*[width=3.in]{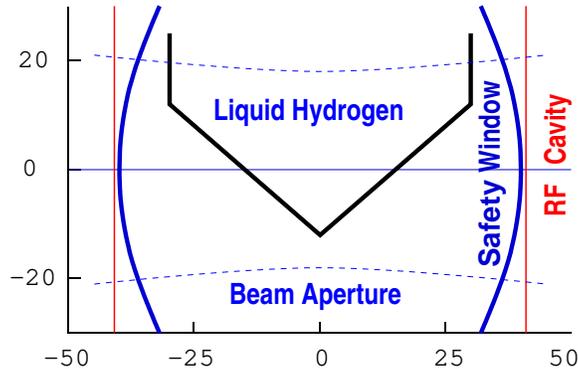}
\end{center}
\caption{(Color) Wedge absorber used in some of the simulations.
} 
\label{bfig5}
\end{figure}

   The liquid hydrogen wedge absorbers have a \textit{house} shape, as shown in Fig.~\ref{bfig5}, and are located in dispersive regions in order to decrease the momentum spread in the beam.

The mean energy loss per cell in the absorber was 8.4 MeV. This energy loss
was exactly compensated by the rf cavities. The energy gain per cell is
\begin{equation}
\Delta E_{rf}=N_{cav}\hat{G}L_{cav}T\sin{\phi_s}.
\end{equation}
where $N_{cav}$ is the number of rf cavities per cell, $L_{cav}$ is the
length of each cavity, $\hat{G}$ is the peak rf gradient, $T$ is
the transit factor and ${\phi_s}$ is the synchronous phase.

 The simulated absorber windows
were planar and located axially just in front of and behind the wedge
itself. In reality the window shape will conform to the shape of the
absorber and the effect on the beam of scattering in the window should be
lessened.  Because of the small bending field we use a wedge with maximum
possible opening angle, which has zero thickness on one side. The wedge has a
central thickness of 28~cm, a total wedge opening angle of $100^{\circ}$ and 
is rotated $30^{\circ}$ from the vertical to match the maximum of the dispersion. This
combination of dispersion and wedge maximizes 6D emittance reduction.

\section{Lattice Functions and Beam Dynamics\label{sec3}}
For our initial studies of the basic lattice functions and beam dynamics we consider the ring with the magnetic field present, but no rf or absorbers. This enables us to simply study some of the basic features of the lattice, such as acceptances, beta function, closed orbits and dispersion functions. 
\subsection{Acceptance}
\begin{figure}[!tbhp] 
\begin{center} 
\includegraphics[width=3.in]{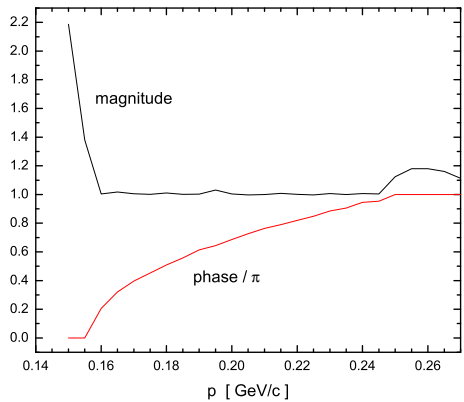}               
\caption{(Color) Magnitude and phase of the largest eigenvalue for the 
one-cell transport matrix} 
\label{bfig6}
\end{center}
\end{figure}

We first examine the acceptance of the magnetic lattice as a function of momentum. A series of particles with small deviations from the closed orbit were tracked through one cell of the ring lattice. The eigenvalues were then extracted from the resulting linear transport matrix. The magnitude and phase of the largest eigenvalue are shown in Fig.~\ref{bfig6}.
 The magnitude is very close to 1 over the momentum band from 160 to 245
 MeV/c. The ring has a very weak linear difference coupling resonance
 excited by the bend of the channel~\cite{rfofo-ref11}. In practice this
 resonance is suppressed by the cooling.

We examined~\cite{rfofo-ref6} the dynamic acceptance of the lattice, with
no rf or absorber, for three different bending fields: 0.0~T, 0.125~T, and
0.25~T. Using ICOOL, particles were injected into the lattice at 9~different
momenta and 6~different angles. 
 We found that the
acceptance was reduced as the bending field was increased. We conclude 
that it is best to use the least bending field consistent with adequate
emittance exchange.
\subsection{Beta function}
\begin{figure}[!tbhp] 
\begin{center}
a) 
\includegraphics[width=3.in]{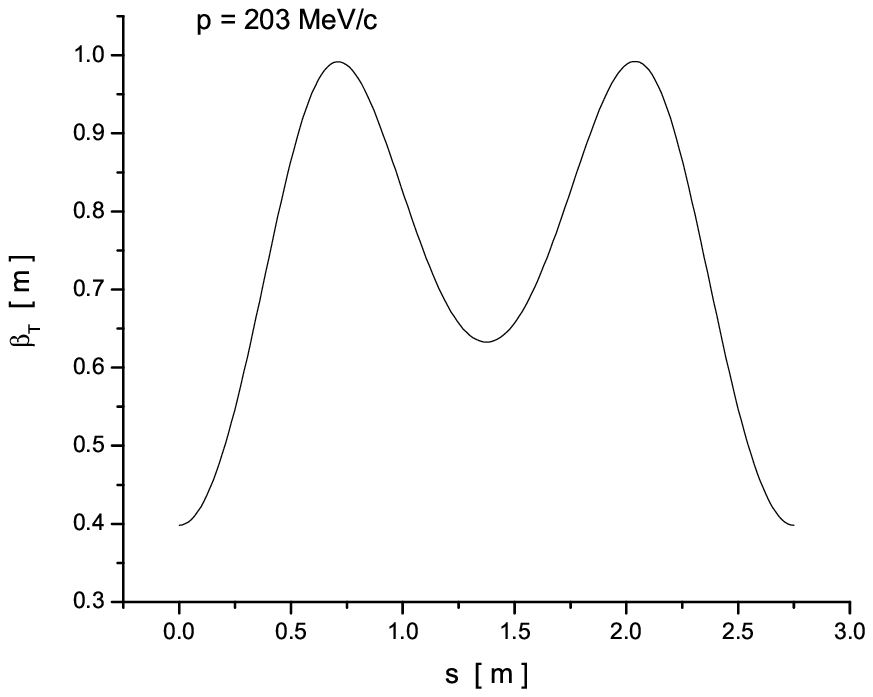}               
b)
\includegraphics[width=3.in]{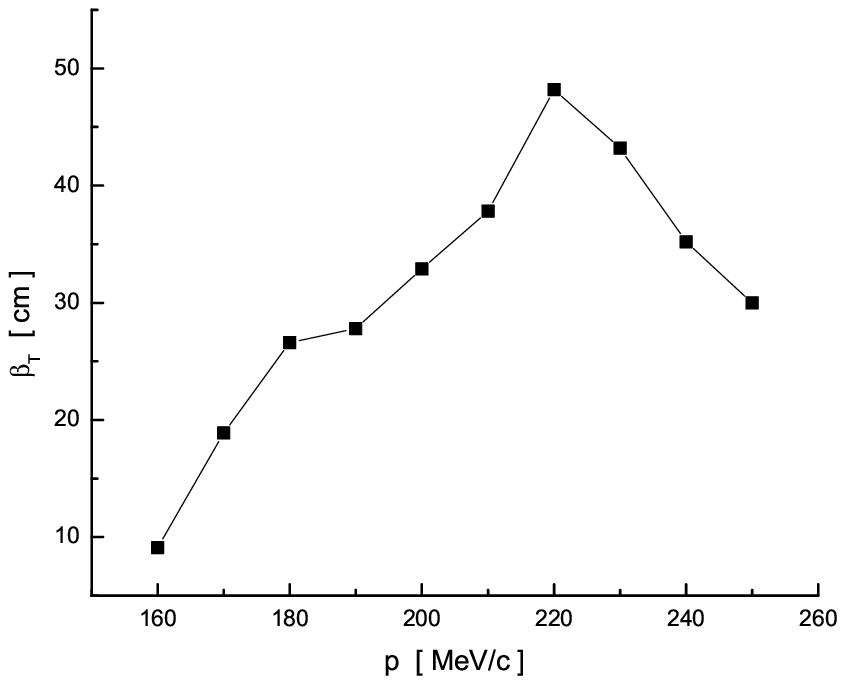}               
\caption{a) Beta function vs. position in the cell; b) beta function vs. muon momentum at the center of the absorber.} 
\label{bfig8}
\end{center}
\end{figure}

The left hand side of Fig.~\ref{bfig8} shows the beta function as a
function of position along a cell. The beta function was determined by
tracking a set of particles with small deviations from the closed orbit
through one cell of the magnetic lattice. Coupling between the transverse
directions was ignored. The minimum value of the beta function at the
center of the absorber ($s=0$ in Fig.~\ref{bfig8}) and at the central momentum is about 38~cm.

The right hand side of Fig.~\ref{bfig8} shows the beta function as a
function of the muon beam momentum. This also shows that the lattice
transmits particles in the momentum band 160--250~MeV/c. These limits are
determined by very strong $\pi$ and $2\pi$  resonances at the limiting momenta.
\subsection{Closed orbits}
\begin{figure}[!tbhp] 
\begin{center} 
\includegraphics[width=3.in]{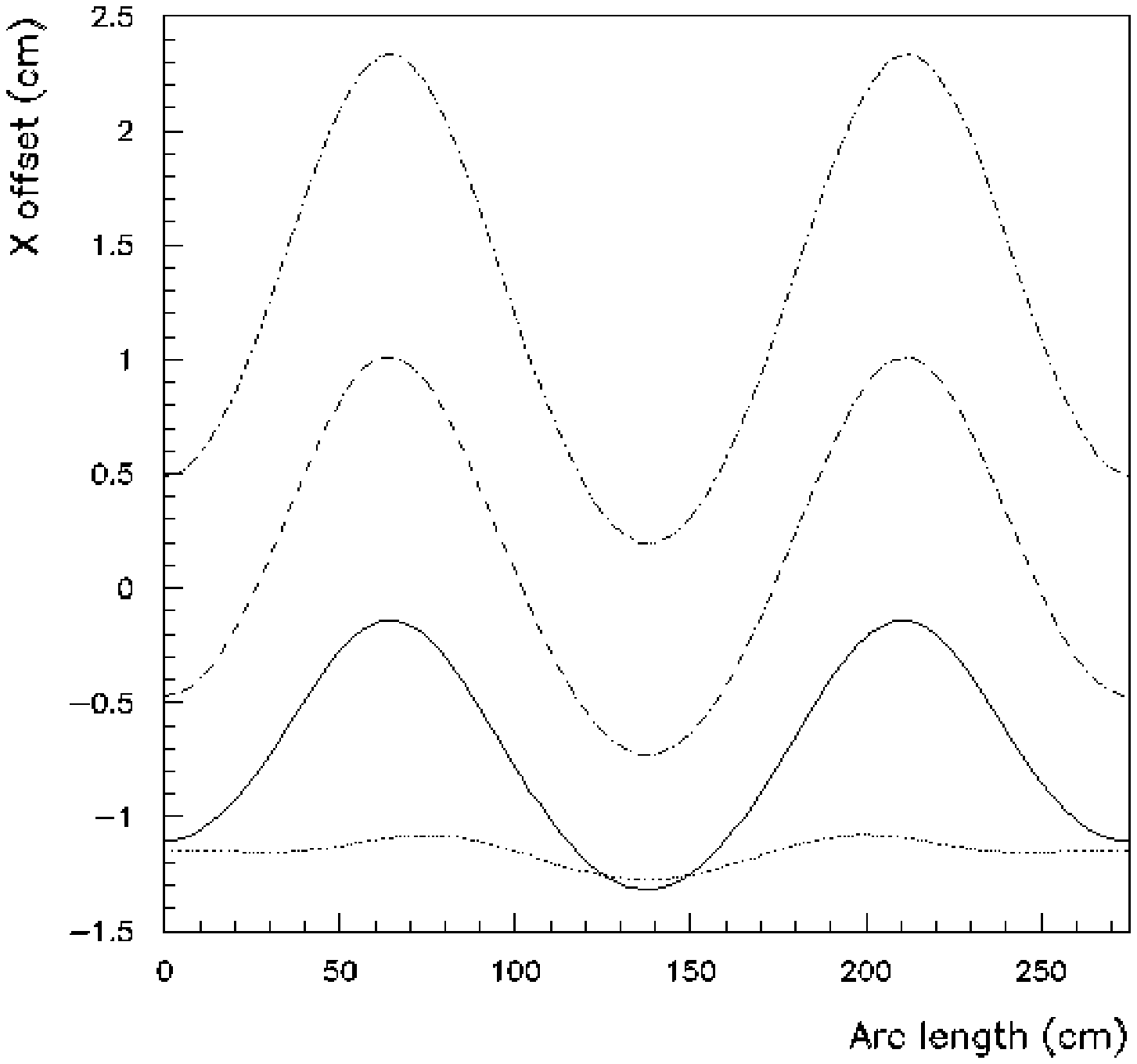}               
\includegraphics[width=3.in]{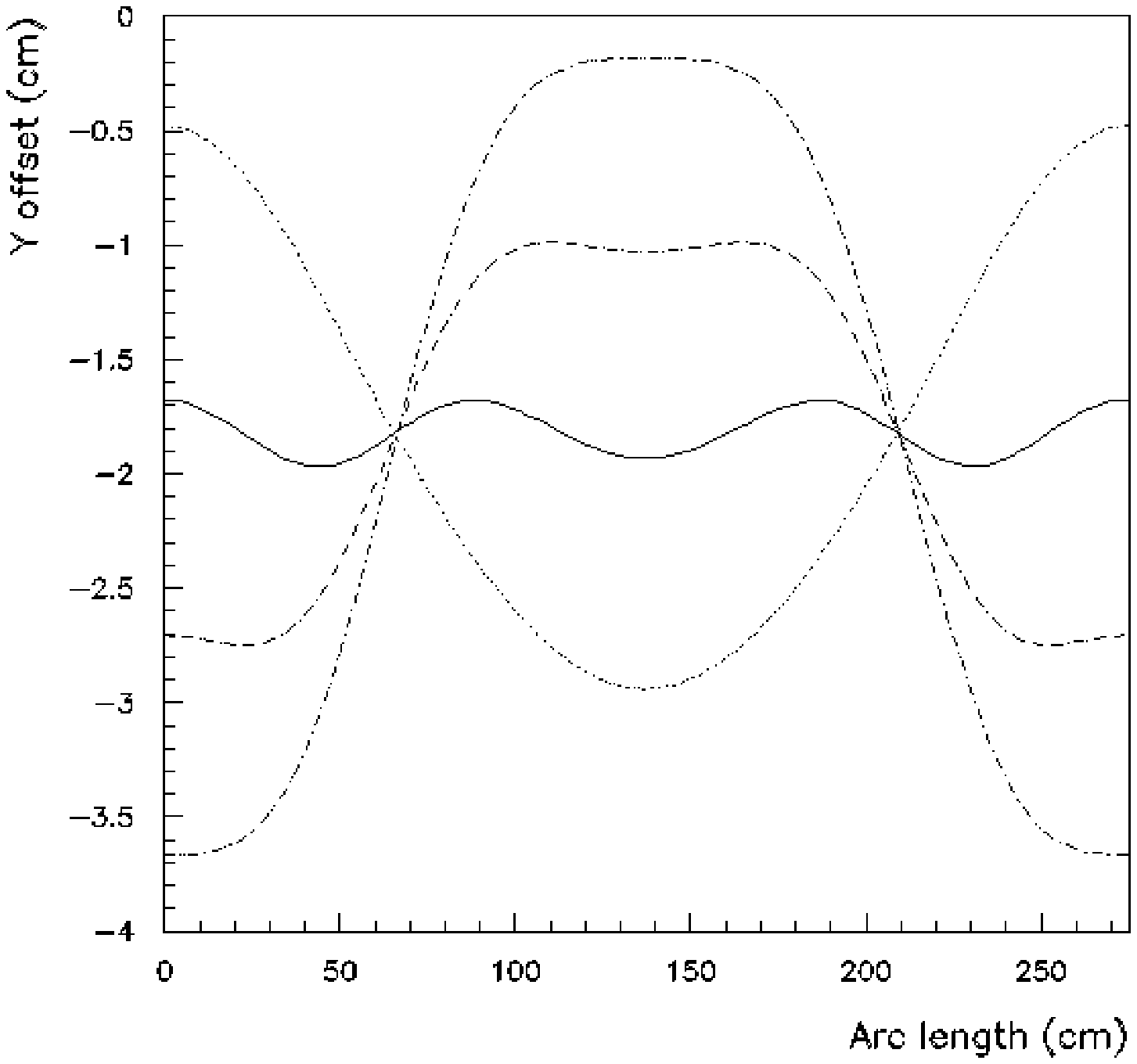}               
\caption{Closed orbits in a cell--deviations from the central line of the
  ring in the radial direction (left) and in the vertical direction
  (right). The different line types correspond to orbits of different
  momentum muons: solid is for 201~MeV/c (\textit{reference particle}),
  dotted, dashed, and dot-dashed are for 170, 227, 248~MeV/c, respectively.}
\label{bfig10}
\end{center}
\end{figure}

Because of the symmetry of the magnetic field, radial and vertical
deviations of the closed orbit must be even functions of s at any
energy. This requires that their derivatives must be 0 at the ends of the
period. This makes it easier to search for closed orbits because only
variations in $x_0$ and $y_0$ need be considered. Muons were
\textit{injected} in the
simulations at the boundary of the cell (the absorber central plane), where
 the transverse momentum
vanishes for a closed orbit. For each initial (longitudinal) momentum, a
unique closed orbit was determined by scanning the plane, and finding the
point to which the muon returns every time it recrosses the absorber
central plane. Since the magnetic field has a small radial component
on-axis, the closed orbits are non-planar. The solid line in
Fig.~\ref{bfig10} shows the transverse motion of
the closed orbit for a 201~MeV/c muon.
 Note that the closed orbit is offset by about 11~mm in $x$ and by about 17~mm in $y$ at
the beginning of the cell. Along the cell $x$ varies by $\pm 6$~mm and $y$
varies by $\pm 2$~mm. The variation of the closed orbits for
changes in momentum are also shown in Fig.~\ref{bfig10}. 
\subsection{Dispersion function}
\begin{figure}[!tbhp] 
\begin{center} 
\includegraphics[width=3.in]{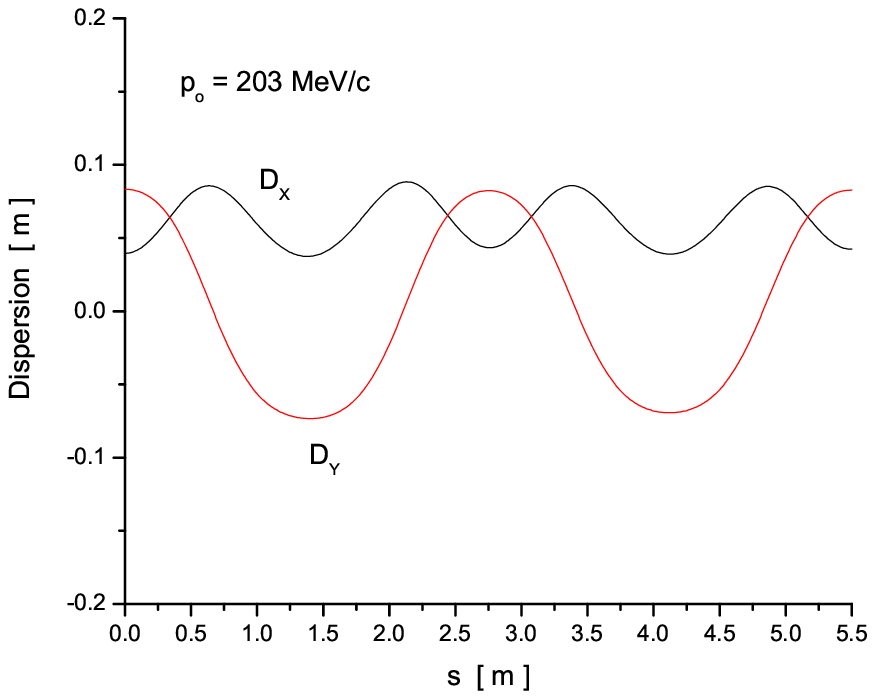}               
\includegraphics[width=3.in]{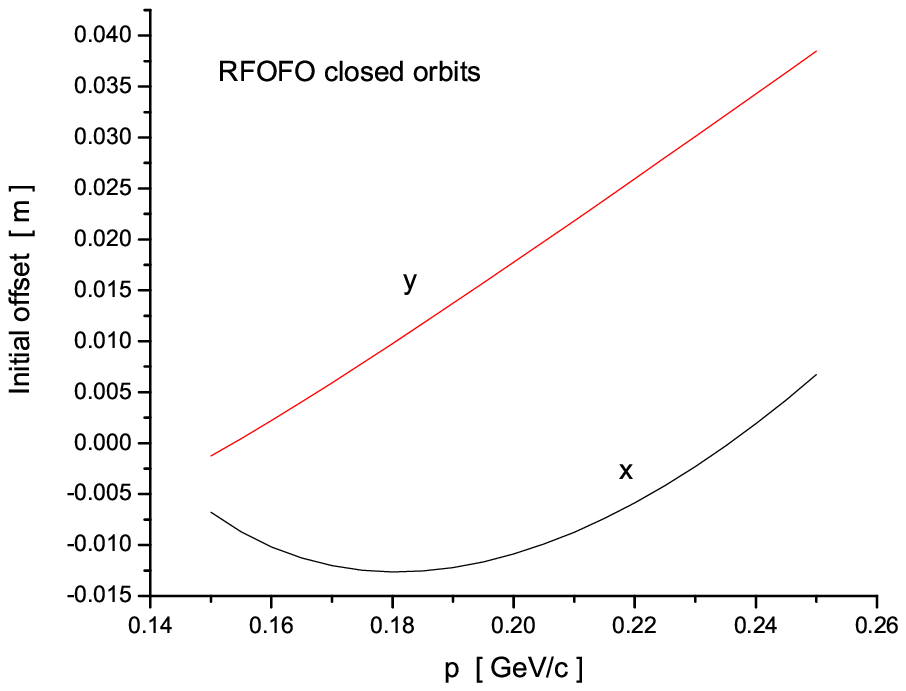}
\caption{(Color) (left) Dispersion vs. position in the cell; (right) initial offset vs. muon momentum.} 
\label{bfig11}
\end{center}
\end{figure}

Dispersion functions in $x$ and $y$ at the absorber central plane (and approximately the
absorber as a whole) can be derived from the closed orbits. One can see the
linear dependence of the y (vertical) offset on the energy in
Fig.~\ref{bfig11}. The dispersion at the absorber is approximately 8~cm in
a direction $30^{\circ}$ from the $y$ axis.
The dispersion at the center of the rf cavity has the opposite sign, and is
also mostly in the $y$ direction. The vertical dispersion is surprisingly
linear as a function of momentum, but the radial dispersion is not.

\section{Cooling Simulations: Ideal Ring\label{sec4}}
We now turn to simulations of muon beams in the ring. For cooling studies
we must now include the rf cavities and absorbers in the simulations.

Several merit factors have been defined for quantifying cooling performance.
\begin{equation}
M(s)=\frac{\epsilon_{6}^N(0)}{\epsilon_6^N(s)}\frac{N(s)}{N(0)},\quad D(s)=\frac{n(s)}{n(0)},\quad
Q(s)=\frac{\frac{d\epsilon_6^N}{ds}}{\frac{dN}{ds}}\frac{N(s)}{\epsilon_6^N(s)}
\end{equation}
For the M-factor $\epsilon_{6N}$ is the normalized 6D emittance and $N(s)$ is the
total number of surviving muons at location $s.$ For the D-factor, $n(s)$
is the number of muons that are contained in a fixed transverse and
longitudinal phase space acceptance volume $V$ after traversing a distance
$s.$ The M-factor is more relevant for a muon collider, while the D-factor
would be more applicable to a neutrino factory. The Q-factor is a local 
variable~\cite{rfofo-ref23} that compares the rate of change of the
emittance to the particle loss.

We first examine the performance of an \textit{ideal} ring, ignoring for
the moment additional effects due to windows on the absorbers and rf
cavities and leaving empty space for injection. Several important ring
parameters are listed in Table~\ref{tb2}. 
\begin{table}[!bthp]
\begin{center}
\caption{Ring parameters}
\label{tb2}
\begin{ruledtabular}
\begin{tabular}{lc}
Curvature $\rho$ (m) & 5.252\\
Momentum compaction $\alpha_P$ & $1.52\, × 10^{-2}$\\
Transition energy factor $\gamma_T$ & 8.09\\
Slip factor $\eta$ & -0.198\\
Longitudinal beta function $\beta_S$ (m) & 6.06
\end{tabular}
\end{ruledtabular}
\end{center}
\end{table}
 $\rho$ is the mean radius of curvature. The momentum
compaction $\alpha_P$ is the fractional change in circumference per fractional
change in momentum. This ring always operates below the transition energy. The
longitudinal beta function $\beta_S$ is given by~\cite{rfofo-ref24}
\begin{equation}
\beta_S^2=\frac{-c\eta CE}{2\pi \beta f eV \cos{\phi_S}}
\end{equation}
 where $C$ is the circumference of the ring, $E$ is the energy of the
 muons, $\beta$ is the relativistic velocity, $e$ is the unit of charge, $V$ is
 the peak rf voltage and $\phi_S$ is the synchronous phase. Note that $\beta_S$ is
 much larger than the transverse beta function $\beta_T.$

A reference particle was used to determine each cavity's relative
phase. The reference particle is a muon that runs in a closed orbit and is
used as a \textit{clock} to determine the rf frequency and set the entry
times for each cavity in the ring. This clock satisfies the following 
conditions:
\begin{itemize}
\item the revolution period is an integer multiple of the rf period
\item the muon momentum is around 200~MeV/c.
\end{itemize}
In GEANT the closed-orbit muon that was chosen as the reference particle
has a momentum of 200.96~MeV/c. With a revolution period of 124.22~ns, its
$25^{th}$ harmonic is 201.26~MHz. For the ICOOL runs we used a circular
reference orbit. 

\begin{figure}[!tbhp] 
\begin{center} 
\includegraphics[width=3.in]{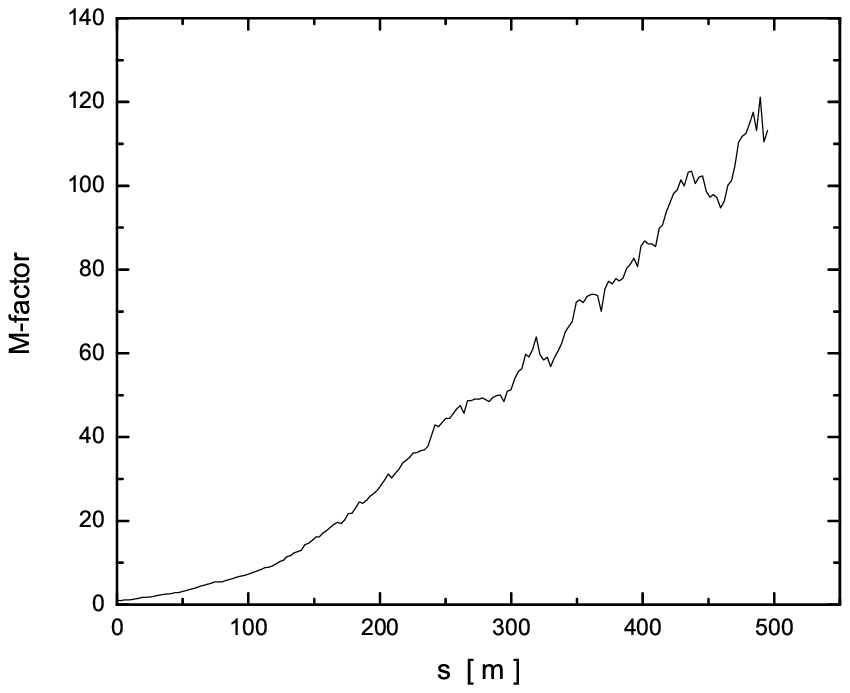}               
\includegraphics[width=3.in]{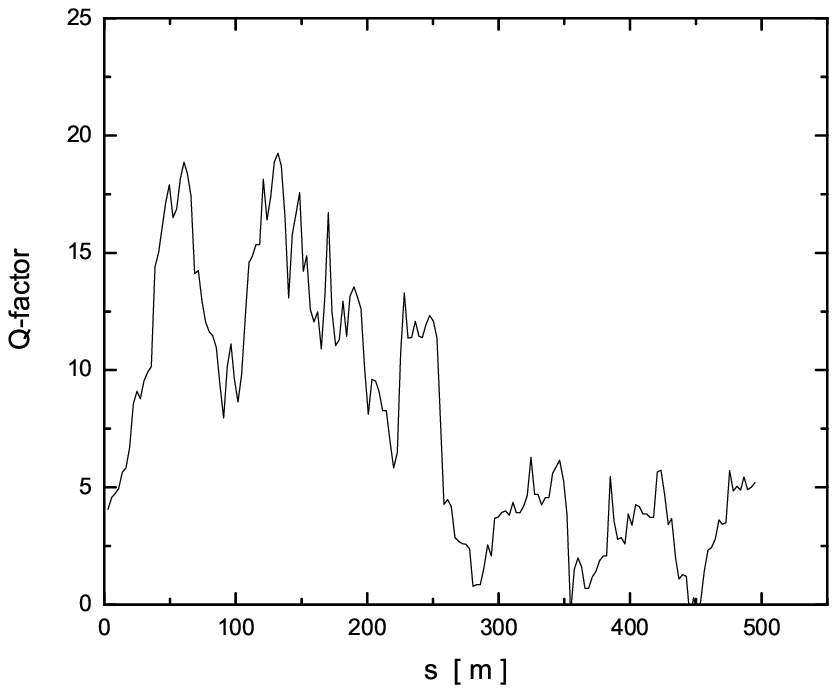}
\caption{M-factor (left) and Q-factor (right) as a function of distance for the ideal ring. 
} 
\label{bfig12}
\end{center}
\end{figure}

Figure~\ref{bfig12} shows the performance of the ideal ring as a function
of distance. For the simulations we use a Gaussian input beam with
normalized transverse emittance of 12~mm and normalized longitudinal
emittance of 18~mm.  The initial beam had a correlation between the axial momentum and the
transverse amplitude to minimize the tendency for the particles in the
bunch to spread out longitudinally in the solenoidal field. The correlation
causes the average axial momentum to be larger than the reference momentum
of 203~MeV/c. 

A separate analysis that decoupled the emittance planes gave final
normalized emittances $\epsilon_{x}^N=2.32$~mm, $\epsilon_y^N=1.81$~mm, and
$\epsilon_z^N=2.89$~mm. After a distance of 495~m (15~turns) the 6D
emittance has fallen by a factor of ~240 with a transmission of 53\% (66\%
without muon decay) and the M-factor is 120. The same factor for the
U.S. Feasibility Study 2 (FS2)~\cite{fs2} cooling lattice without 
windows is 15.  This ideal ring has a maximum Q-factor of approximately 18.

\begin{figure}[!tbhp] 
\begin{center} 
\includegraphics[width=4.in]{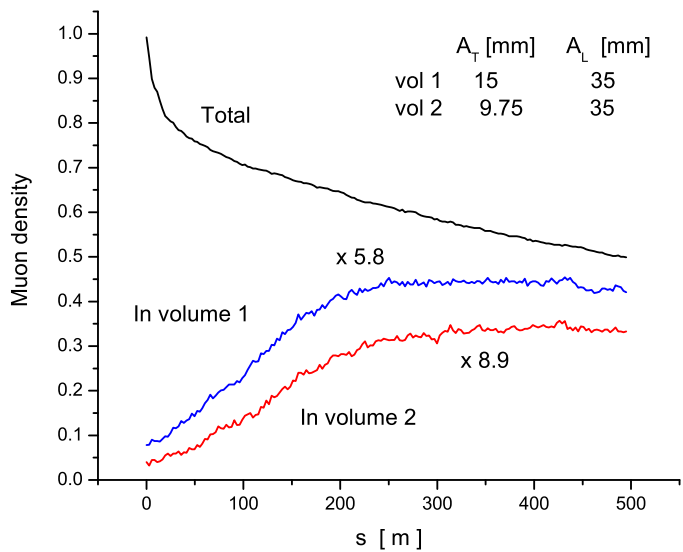}               
\caption{(Color) Performance of RFOFO ring. Transmission and muon density
  into two fixed acceptance volumes. These results were obtained using a
  $100^{\circ}$ wedge.} 
\label{bfig13}
\end{center}
\end{figure}

We next consider the idealized ring behavior in terms of the muon
density. Figure~\ref{bfig13} shows the total muon transmission together
with the muon density into two fixed acceptance volumes. These volumes
correspond to the assumed acceptance of a linear accelerator that follows
the cooling ring.
The idealized ring increases the muon density into the smaller acceptance
volume by a factor of almost 9 in 250~m, which corresponds to about
8~turns. The density in the larger acceptance volume increases by about a 
factor of 6.

\begin{figure}[!tbhp] 
\begin{center} 
\includegraphics[width=3.in]{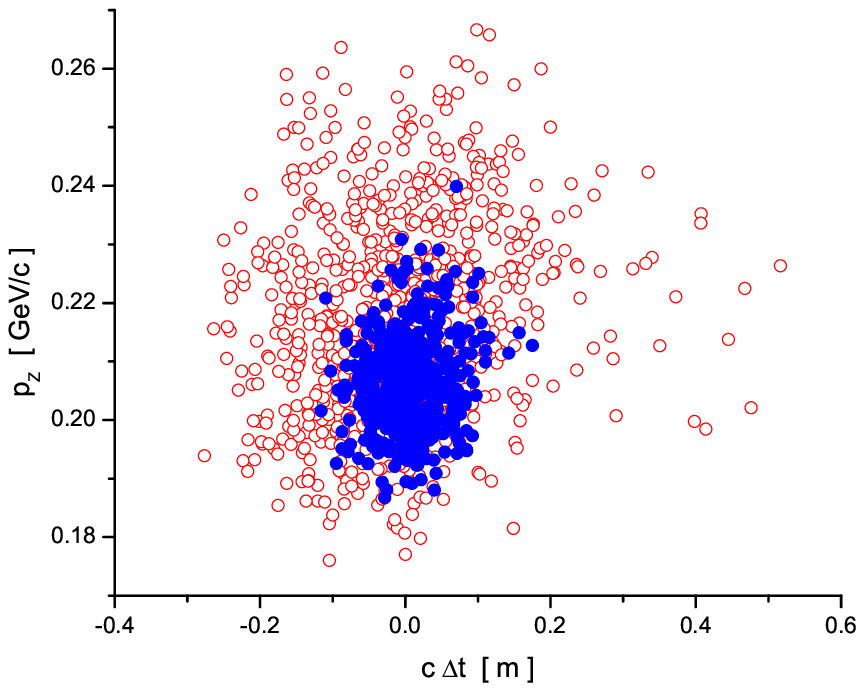}               
\includegraphics[width=3.in]{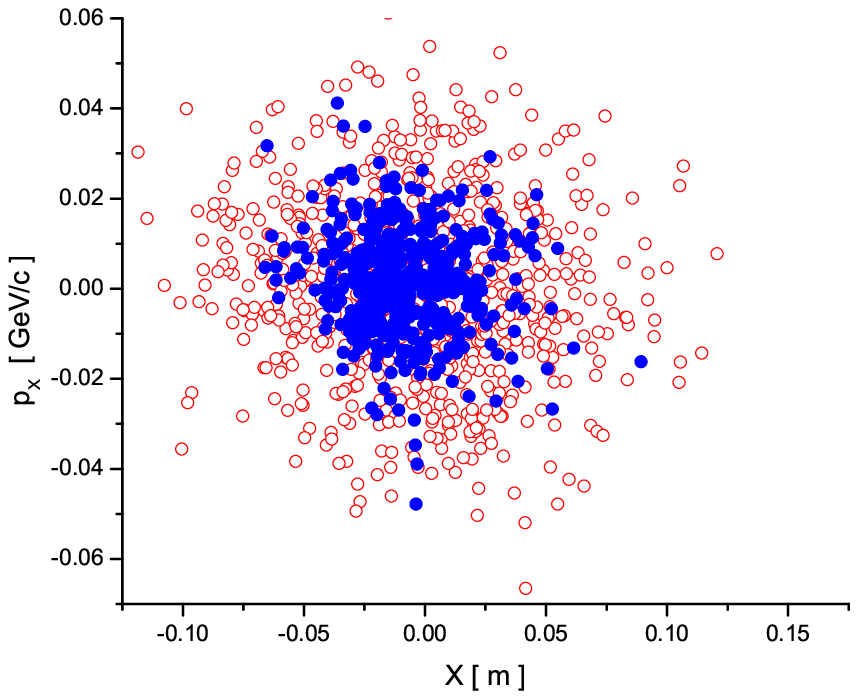}               
\caption{(Color) Longitudinal phase space (left) and radial phase space (right) after 1~turn (open circles) and 15~turns (closed circles).} 
\label{bfig14}
\end{center}
\end{figure}

Figure~\ref{bfig14} shows the radial and longitudinal phase space after 1
and 15~turns.
 The reduction of phase space area can be seen clearly in both distributions. The vertical transverse distribution is similar to the radial one because of the mixing caused by the solenoids. 

\begin{figure}[!tbhp] 
\begin{center} 
\mbox{
\includegraphics*[width=0.4\linewidth]{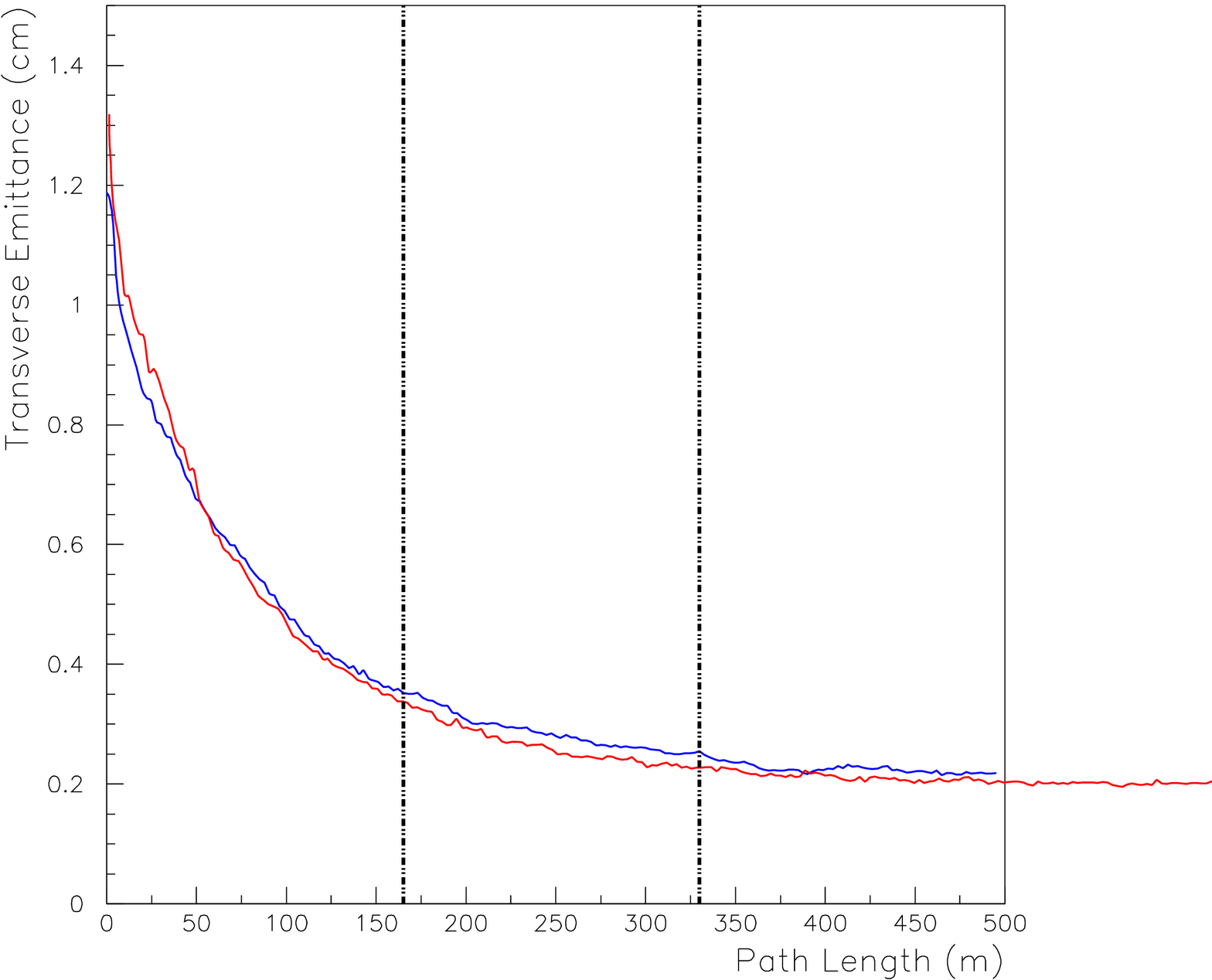}               
\includegraphics*[width=0.4\linewidth]{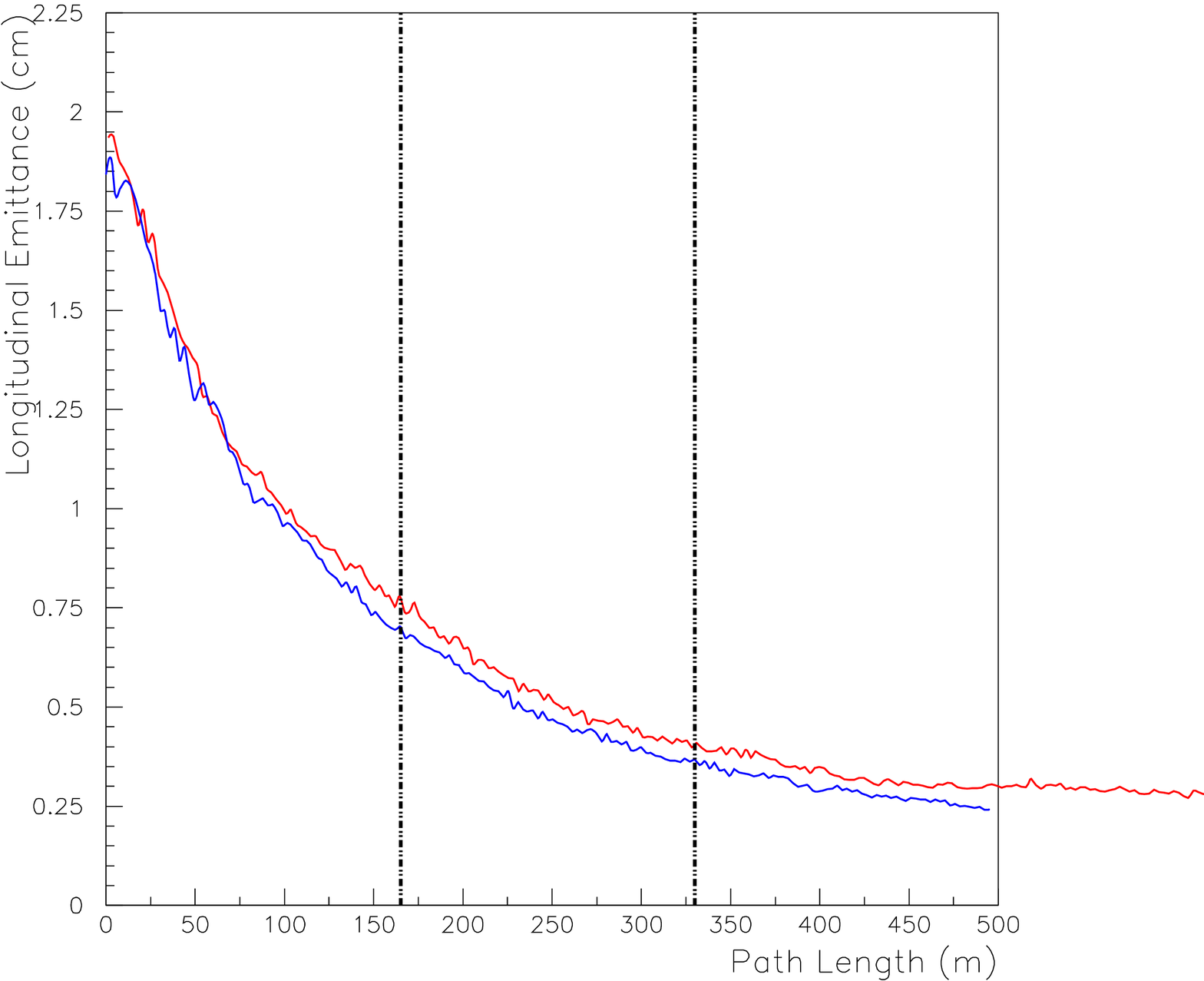}               
}
\mbox{
\includegraphics*[width=0.4\linewidth]{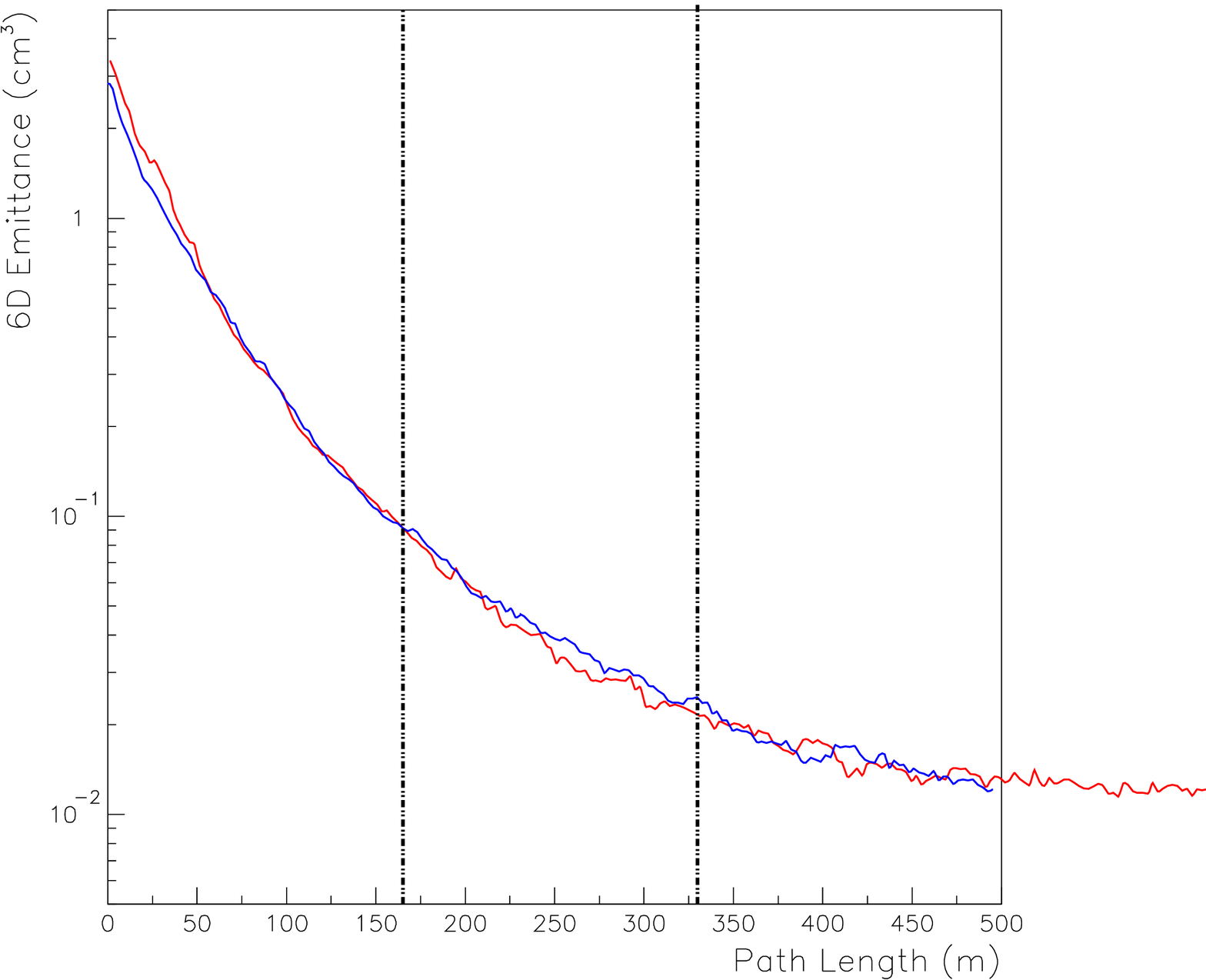}
\includegraphics*[width=0.4\linewidth]{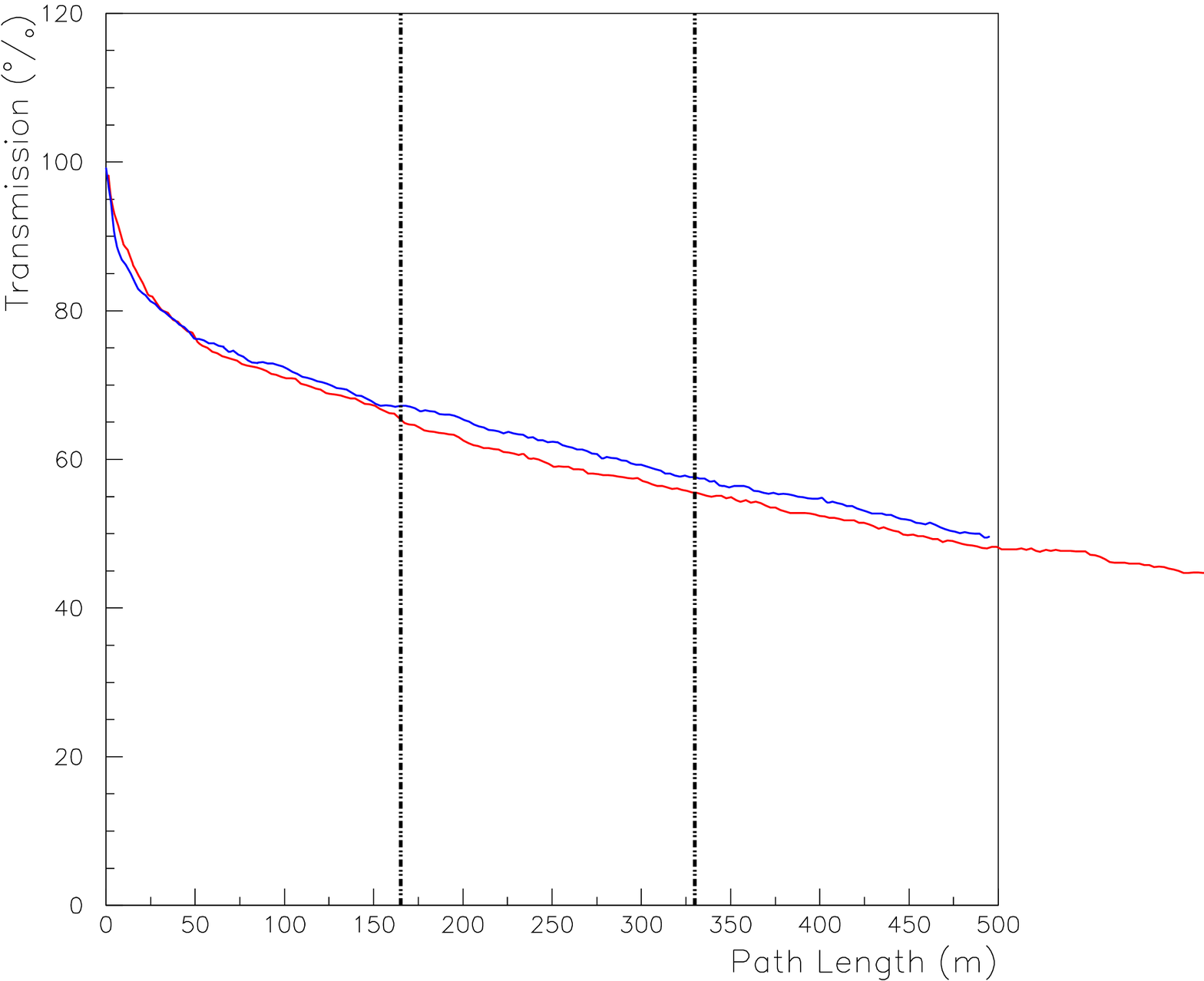}          
}
\includegraphics*[width=0.4\linewidth]{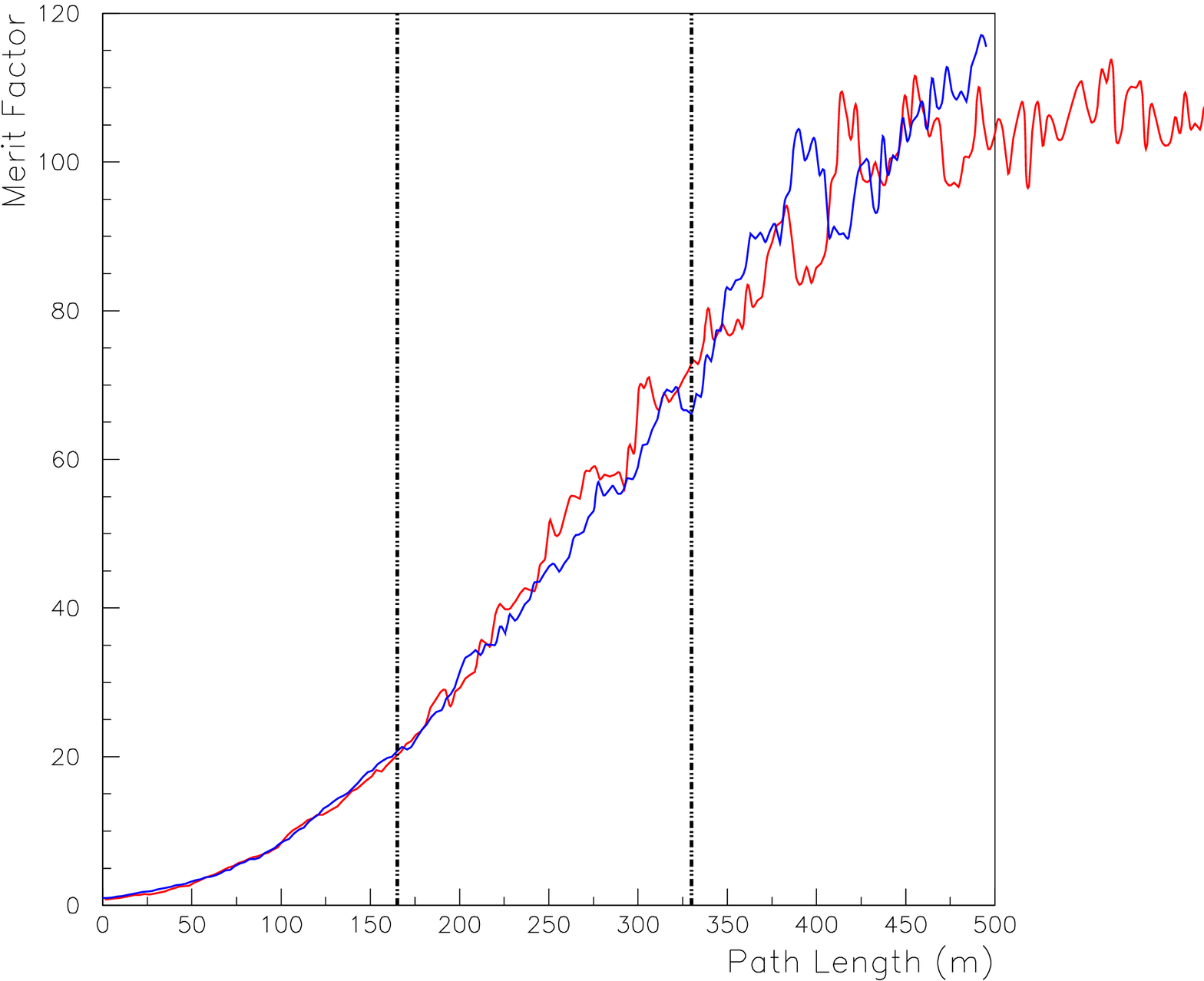}
\caption{(Color) A comparison between GEANT (red line) and ICOOL (blue)
  for: (top panel) transverse emittance and longitudinal emittance; (middle
  panel) 6D emittance
  and transmission; (lower panel) M-factor vs. path length. The vertical lines indicate 5 and 10 ring turns  respectively.} 
\label{bfig15}
\end{center}
\end{figure}

A comparison of ICOOL and GEANT performance calculations is shown in Fig.~\ref{bfig15}. The initial drop in transmission over the first few turns
includes a contribution from scraping at transverse apertures due to
mismatches between the initial beam distributions and the ring
acceptance. The emittance was calculated using the program ECALC9~\cite{rfofo-ref21,rfofo-ref21a}, which is a standard tool used for this purpose.
The agreement in calculated performance between the two codes is excellent.
\section{Cooling Simulations: Realistic Ring\label{sec5}}
We now consider the effects on performance of including windows for the
absorbers and rf cavities and leaving empty cells for injection. We refer
to this as the \textit{realistic} ring design. Including windows
introduces new sources of scattering and degrades the cooling
performance. We enclose the liquid hydrogen absorbers with aluminum
windows. The rf cavities have tapered beryllium windows.  In general the
M-factor is more sensitive to the presence of windows than the
D-factor. Two adjacent cells have the absorber and rf cavities removed in
order to leave space for the injection kicker, but the magnetic periodicity
is maintained. We show the influence of these effects on the peak value of
the D-factor in Table~\ref{tb3}. 

The absorber window used in FS2 was $360\,\mu \text{m}$ of aluminum. We see
that windows of this thickness degrade the performance by about 30\%. For
safety reasons it may be necessary to use an additional window that
increases the total amount of aluminum. On the other hand the use of other
special materials or optimized window shapes could reduce the amount of
material required. Another possibility would be to replace the liquid H$_2$
absorber with a solid material. LiH is one possible candidate, although
Table~\ref{tb3} shows there is a 45\% loss of performance with this option.
\begin{table}[!bthp]
\begin{center}
\caption{Perturbations on the ideal ring performance. FS2 stands for
  windows used in Study 2; FS2/4 stands for windows used in 
FS2 with thicknesses reduced by a factor of 4.}
\label{tb3}
\begin{ruledtabular}
\begin{tabular}{lcccc}
Absorber&Abs. Window &rf Window &Empty cells &D-factor\\
LH&none&none&0&8.93\\
LH&$250\,\mu\text{m}$ Al &none &0&7.50\\
LH&$360\,\mu\text{m}$ Al &none &0&6.70\\
LH&$500\,\mu\text{m}$ Al &none &0&6.08\\
LiH &none & none &0&4.88\\
LH &none & FS2 &0&5.88\\
LH &none & FS2/4 &0&7.80\\
LH &none & none &2&6.73\\
LH&$360\,\mu\text{m}$ Al &FS2/4 &2&4.58
\end{tabular}
\end{ruledtabular}
\end{center}
\end{table}

In FS2 the Be end windows on the rf cavities were $200\,\mu$m
thick from the axis out to a radius of 12~cm, then $400\,\mu$m thick out to
18~cm. The interior windows were $700\,\mu$m thick from the axis to a radius
of 14~cm, then $1400\,\mu$m thick out to 21~cm. These rf windows degrade the
performance by about 35\%. One possibility to get around this problem is to
operate the cavities at liquid nitrogen temperature. The lower operating
temperature and the reduced rf gradient of 12~MV/m versus 16~MV/m in FS2
allow the thickness of the windows to be decreased by a factor of 4 with a
 performance loss of only 13\%. An R\&D program on low-temperature windows
 might permit further reduction in the rf window thicknesses. Alternatively, one could eliminate the rf
cavity windows altogether and use an open cavity. This has the disadvantage
that four times more power is required to produce the same $E_Z$ on axis,
assuming that surface breakdown does not limit the axial field available.

\begin{figure}[!tbhp] 
\begin{center} 
\includegraphics[width=4.in]{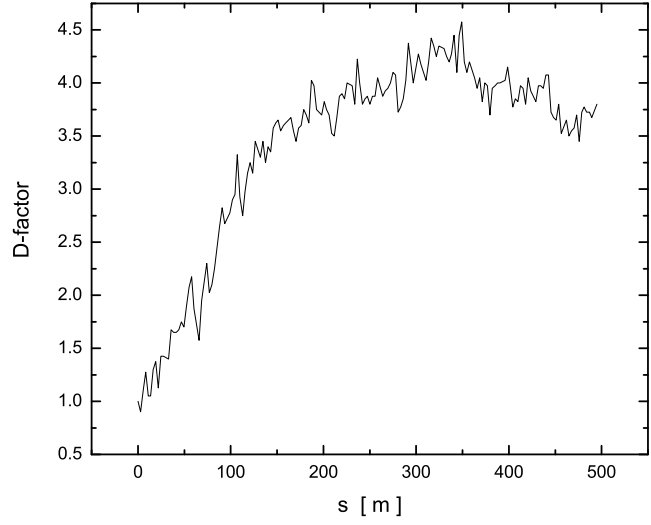}               
\caption{D-factor vs. distance for the realistic ring.} 
\label{bfig16}
\end{center}
\end{figure}

Introducing two empty cells for injection/extraction reduces the performance by 25\%. Finally, we consider an example that combines all of these effects. We choose liquid H$_2$ as the absorber with $360\,\mu$m thick windows and no safety windows. We assume that operation at liquid nitrogen temperature allows the thinner Be rf windows, and leave the empty cells in the lattice for injection/extraction. Figure~\ref{bfig16} shows the evolution of the D-factor for the realistic ring.
The initial phase space volume used in this simulation had a normalized
transverse acceptance of 9.75~mm and a normalized longitudinal acceptance
of 35~mm. The muon density into the accelerator acceptance peaks after
about 10~turns. This realistic ring model still gives an impressive
increase in the accepted muon density of a factor of $\approx 4.5.$

\begin{figure}[!tbhp] 
\begin{center} 
\includegraphics[width=4.in]{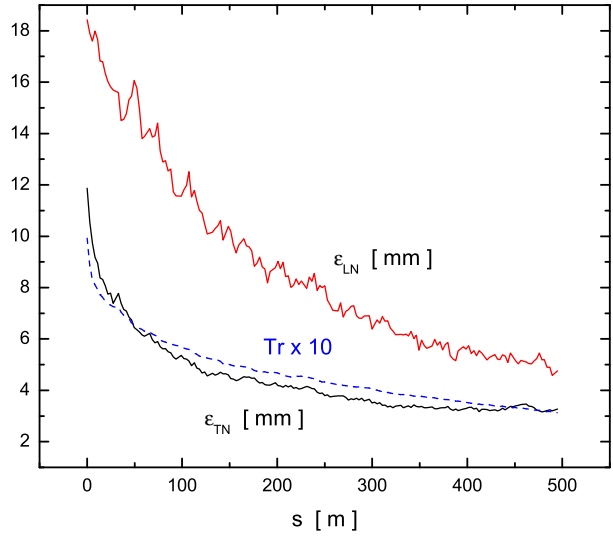}               
\caption{(Color) Normalized emittances and transmission with decay included
  as a function of accumulated distance in the realistic cooling ring. One
  turn is 33~m.} 
\label{bfig17}
\end{center}
\end{figure}

Figure~\ref{bfig17} shows the evolution of the emittances and transmission for the realistic ring design.
Some of the transverse losses originate from particles that initially slip
out of the rf bucket. Following this there is a steady loss due to
decays. Initially the $x$ emittance falls more rapidly than the $y$ because
it is the $\epsilon_y^N$ emittance that is mainly exchanged with the longitudinal. But
the Larmor rotations in the solenoids soon mix the $x$ and $y$ emittances,
bringing them close to a common value. The emittance performance of the realistic
ring is summarized in Table~\ref{tb4}.
\begin{table}[!bthp]
\begin{center}
\caption{Emittance performance of the realistic RFOFO cooling ring.}
\label{tb4}
\begin{ruledtabular}
\begin{tabular}{lcc}
&Initial&Final\\
$\epsilon_T^N$ (mm) &11.9&3.3\\
$\epsilon_L^N$ (mm) &18.4&4.8\\
$\epsilon_6^N$ (mm$^3$) &2830&58\\
Tr (\%) & & 31\\
M & &15
\end{tabular}
\end{ruledtabular}
\end{center}
\end{table}

The mean angle of muons at the end of the channel was 66~mr and the maximum angle for an accepted particle was 200~mr. This illustrates the advantage of using solenoidal focusing for this application. Although the realistic ring merit factors are reduced considerably from the ideal ring, where the corresponding merit factors were $M=120$ and $D=8.9,$ the cooling performance is still quite significant.
\section{Design Issues\label{sec6}}
There are a number of difficult design issues that must be resolved before
a real ring cooler can be built. Each of the following areas needs a
program of experimental R\&D.
\subsection{Beam injection and extraction}
\begin{figure}[!tbhp] 
\begin{center} 
\includegraphics[width=3.in]{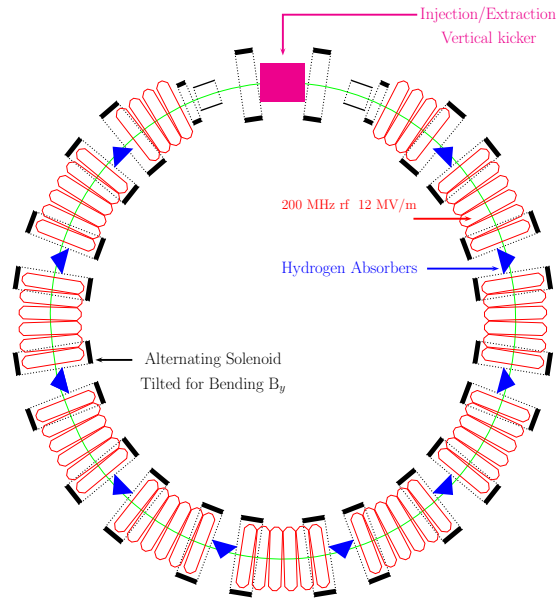}               
\caption{(Color) Layout of RFOFO ring with injection/extraction cells} 
\label{bfig18}
\end{center}
\end{figure}

The design of the injection-extraction channels and kickers will be challenging. The 12-fold symmetry of the ring 
must be broken to allow space for injection and extraction. In some simulations two cells have been modified for 
this purpose, as shown in Fig.~\ref{bfig18}. 

\begin{figure}[!tbhp] 
\begin{center} 
\includegraphics*[width=5.in]{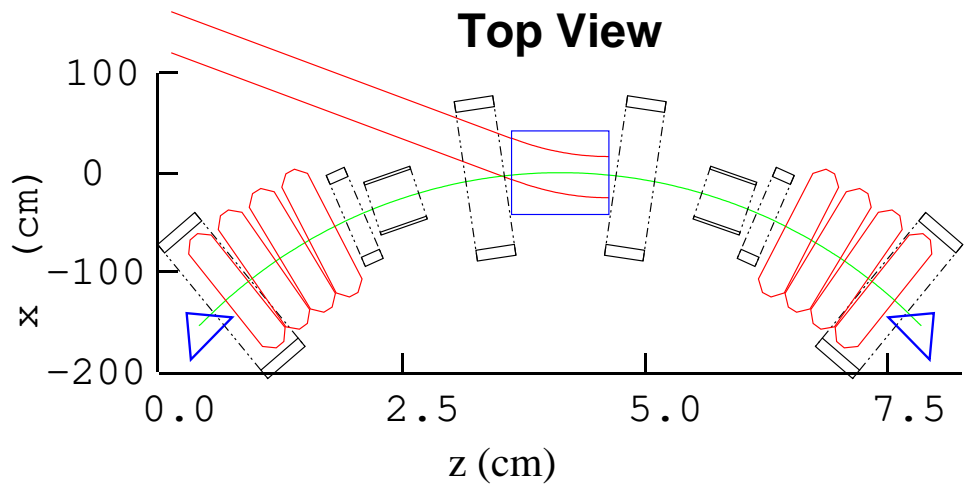}               
\caption{(Color) View of the injection/extraction area in a RFOFO ring} 
\label{bfig19}
\end{center}
\end{figure}

A more detailed layout of the injection area is shown in Fig.~\ref{bfig19}.
 The coils in the insertion area have been modified to allow muons from an external beamline to access the kicker. It is possible to generate magnetic fields in the insertion cells that are nearly identical to those in the rest of the ring. 

The minimum pulsed energy required for the kicker is proportional to the
square of the transverse emittance being kicked. Since the initial muon
emittances discussed here are much larger than those encountered in other
applications, such as antiproton accumulation, the energy in the pulsed
kicker is three orders of magnitude greater, e.g. 10000~J here compared
with 10--20~J for the antiprotons. However, magnetic amplifiers used in
induction accelerators can provide this kind of pulse energy at the
required pulse lengths, and a kicker design based on this concept has been
proposed~\cite{rfofo-ref22}.
\subsection{Absorber heating}
Using a ring for ionization cooling causes the beam bunches to pass through
a given absorber many times. This results in the thermal load on the
absorber in a ring being much larger than for the same absorber used in a
linear channel. Consider the example shown in Table~\ref{tb5}, which uses a
1~MW beam with similar parameters to FS2.
\begin{table}[!bthp]
\begin{center}
\caption{Absorber heating example.}
\label{tb5}
\begin{ruledtabular}
\begin{tabular}{lc}
Muons/bunch &$4.6\,× 10^{12}$\\
Bunch rep. rate (Hz)&15\\
Absorber length (cm)&28.6\\
Turns in the ring& 8\\
Energy deposit/bunch (J)&94 \\
Average beam radius (cm)&4\\
Temperature rise/bunch $({}^{\circ}C)$& 0.21\\
Average power dissipated (kW)&1.42\\
Flow for $\Delta T=2^{\circ}C$ $(ls^{-1})$&1.45\\
Velocity in 5~cm pipe (m/s)&0.74 
\end{tabular}
\end{ruledtabular}
\end{center}
\end{table}
  The average power deposited in the absorber exceeds 1.4~kW. The last two rows give the required water flow and velocity in a typical pipe to extract the heat under steady state conditions. 
\subsection{RF windows}
\begin{figure}[!tbhp] 
\begin{center} 
\includegraphics*[width=5.in]{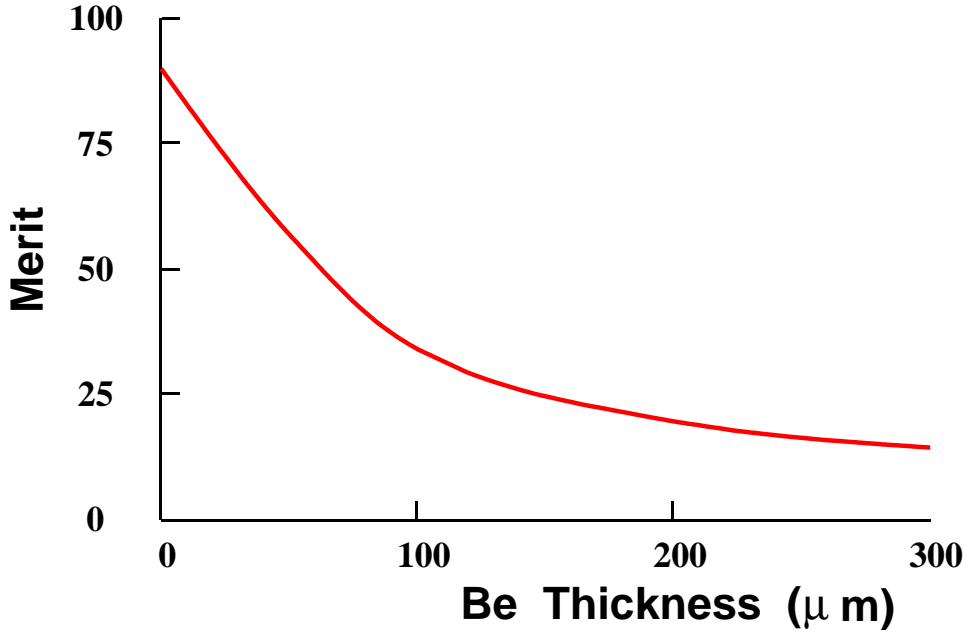}    
\caption{(Color) Dependence of M-factor on rf window thickness} 
\label{bfig20}
\end{center}
\end{figure}

We have seen that the M-factor is very sensitive to the window thicknesses
used in the simulations. Figure~\ref{bfig20} shows how the M-factor for an
early ring design depends on the rf window thickness.
 Thin windows are clearly advantageous, and we would like to use beryllium window thicknesses on the order of 
$25\,\mu m$ if possible. Cooling the cavity to liquid nitrogen temperatures would help.

\section{Design Variations\label{sec7}}
We have also considered several variations of the 33 m circumference, 200 MHz ring discussed above.
\subsection{16.5 m circumference ring}
We have simulated~\cite{rfofo-ref6} a case with twice the bending field
(0.25~T) and thus a ring circumference of only 16.5~m. Such a ring would
require an even shorter initial bunch train, and is not expected to have as
good a dynamic aperture. The transmission of an ideal ring after 15~turns
was 45\%, the transverse normalized emittance was cooled from 10.8 to
3.6~mm, the longitudinal normalized emittance was cooled from 51 to 5.2~mm,
 and the ideal M-factor was 40. The greater dispersion allowed a greater
 reduction in longitudinal emittance at the expense of less transverse
 cooling. The overall merit factor is not as good 
as the standard ring.
\subsection{66 m circumference ring}
A larger circumference ring would allow a longer bunch length of the
initial beam and would reduce problems with the injection kicker. The
average vertical bending field was 0.064~T. The transmission of an ideal
ring after 15 turns was 49\%, the transverse normalized emittance was cooled
 from 11.9 to 2.0~mm, the longitudinal normalized emittance was reduced
 from 19 to 6.1~mm, the M-factor was 60, and the D-factor was 5.2. The
 smaller dispersion hurt the amount of 
longitudinal cooling and lowered the achievable merit factors. 
\subsection{10 MHz ring}
One solution to the problem of a long incoming bunch (train) is to reduce the rf frequency to 10 MHz. Simulations of the 33~m ring were done using a gradient of 2~MV/m and $51^{\circ}$ wedges. The transverse normalized emittance of an ideal ring was cooled from 18.6 to 2.9~mm, the longitudinal emittance was cooled from 251 to 79 mm and the M-factor was 61. The transverse cooling was good, but the longitudinal cooling needs more dispersion. The lower gradient was probably responsible for the slower cooling.
\subsection{Open rf cavities}
It is also possible to use open rf cavities in the ring. This eliminates
the scattering in the rf windows, but requires more rf power to achieve the
same gradient. Simulations with 200~MHz open cavities in an ideal ring
obtained the same performance as closed cavities with $50\,\mu\text{m}$ thick
windows.

\subsection{Cylindrical absorber}
We studied the effect of replacing the \textit{house} shaped absorber with a more practical cylindrical shape. Simulations for an ideal ring using a cylindrical analog to the $100^{\circ}$ wedge gave an M-factor of 109 compared to 152 for the house shape.

\section{Conclusions\label{sec8}}
Cooling large emittance muon beams is important for neutrino factories and
absolutely essential for a muon collider. 
Using a ring cooler geometry
offers likely economic advantages in reusing expensive magnets, rf cavities
and liquid hydrogen absorbers. A ring also provides a natural mechanism for
obtaining longitudinal cooling through emittance exchange. The RFOFO ring
cooler described here is at this time the most advanced ring cooler
design. It uses realistic field modeling and takes into account the effects
of windows on the absorbers and rf cavities and the effects of empty cells
for injection and extraction. 

The ring can increase the muon density in a
standard acceptance volume by over a factor of 4. However, before such a
ring could actually be built, a number of difficult technical issues, such
as the injection kicker, need to be resolved through a program of experimental
R\&D. An intensive R\&D program on rf and absorber windows is presently
underway as part of the International Ionization Cooling 
experiment (MICE)~\cite{rfofo-ref26}.

\begin{acknowledgments}
This research was supported by the U.S. Department of Energy under
Contracts {No. DE-AC02-98CH10886}, {No. DE-AC02-76CH03000} and
U.S. National Science Foundation under Award No. PHY-0104619.
\end{acknowledgments}
\bibliography{rfofo}
\end{document}